\documentstyle[proceedings,numreferences]{crckapb}

\begin{opening}
\title{A NEW TYPE OF MASSIVE SPIN-ONE BOSON:\protect\\
       AND ITS RELATION WITH MAXWELL EQUATIONS}

\author{D. V. AHLUWALIA}
\institute{Los Alamos Meson Physics Facility, P-25\\
           H-846, Los Alamos National Laboratory\\
	   Los Alamos, NM 87545 (USA)\\ \\
E-mail: av@p25hp.lanl.gov}

\end{opening}

\runningtitle{A NEW TYPE OF SPIN-ONE BOSON}

\begin{document}


\section{Introduction}

The text book understanding of quantum field theory states that a fermion and
its associated antifermion have {\it opposite} relative intrinsic parity; and
that a boson and its associated antiboson carry {\it same} relative intrinsic
parity. No particles that do not fall within this understanding, coupled with
the fact that no quantum field theory was known to exist that contradicted
this canonical wisdom, has led to  almost complete neglect of the fact that
long ago Wigner \cite{Wigner}
\footnotemark[1]
\footnotetext[1]{ Part of the
work  in Ref. [1] was done in collaboration with V. Bargmann and A. S.
Wightman, as Wigner \cite{Wigner} points out and we have noted in Ref.
\cite{BWW}. Also similar work was done previously and independently, as we
learned later while writing \cite{DVAn},  by L. L. Foldy and B. P. Nigam
\cite{Foldy,Nigam}. So, whenever we refer to Wigner-type bosons or theory, the
reader may wish to keep this historical note in mind. However, it must be
noted that the work of Ref. \cite{Wigner} is the most comprehensive and is
within
a much deeper and fundamental framework. {\sl I thank Zurab K. Silagadze for
kindly bringing to my attention the work of Foldy and Nigam on the subject
\cite{ZKS}.
Also see Acknowledgements.}}  argued that space-time symmetries, and that
these manifest in quantum constructs as projective representations, do allow
for theories where  a fermion and its associated antifermion have {\it same}
relative intrinsic parity; and that a boson and its associated
antiboson carry {\it opposite} relative intrinsic parity. The only scholarly
footnote to the canonical wisdom in the widely read literature, see for
example Refs. \cite{QFT},  that refers to Wigner's observations is
Appendix C of Weinberg's latest monograph on the quantum theory of fields
\cite{SWqf}.

Recently \cite{BWW}, purely by accident (see Acknowledgements),
in an attempt to understand an
old work of Weinberg \cite{SW} and to investigate the possible kinematical
origin for the violation of P, CP, and other discrete symmetries
\cite{DVAn},  a Wigner-type quantum field theory was constructed for a
spin-one boson. This new theory made it clear that {\it  boson-antiboson
relative intrinsic parity is not (necessarily and) uniquely determined by
the representation space in which the theory is constructed.}  For example,
 the $(1/2,1/2)$ representation space describes spin-one particles in which
a particle and its antiparticle  carry {\it same} relative intrinsic
parity. \footnotemark[2]\footnotetext[2]{At this time (September 1995) it
is not known if $(1/2,1/2)$ representation space too can support a theory of
the Wigner type.}  On the other hand \cite{BWW}, the $(1,0)\oplus(0,1)$
representation space not only allows for the construction of a theory in
which boson-antiboson relative intrinsic parity is same (``Weinberg's
theory'' \cite{SW}) but also supports a construct in which boson-antiboson
relative intrinsic parity is opposite (``our Wigner-type theory''
\cite{BWW}). This is a surprising result, particularly in view of the
canonical wisdom that states that there is no physical distinction between
the two representation spaces,
i.e., $(1/2,1/2)$ and $(1,0)\oplus(0,1)$,
 for the description of spin-one particles.
\footnotemark[3]
\footnotetext[3]{The canonical wisdom translated to the spin-one case under
consideration can now be stated as:
There is no physical distinction between  the usual
(See footnote[2])
$(1/2,1/2)$ representation space description
and the Weinberg's construct \cite{SW} in the $(1,0)\oplus(0,1)$
representation space.}
For an eloquent and forceful argument putting forward the canonical wisdom,
one may refer to Weinberg's paper \cite{SWws} presented at the fiftieth
anniversary of the famous 1939 Wigner's paper \cite{EPWf} on the unitary
representations of the inhomogeneous Lorentz group. In fact Weinberg
\cite{SWpc} himself was surprised at the new theory and wrote ``When I saw
your paper I did not believe the result. ... I suspect that this is why you
are getting such a surprising result. In all my work, I assume that states
of given mass and spin are non-degenerate.''

The Wigner-type quantum theory of fields is a very natural consequence of
space-time symmetries and the fundamental structure of quantum mechanics.
The underlying physical principles for the standard quantum theory of
fields and Wigner-type theory are the same --- standard quantum theory is one
of the classes of the Wigner's general classification \cite{Wigner}.
Our recent construction
of  a Wigner-type theory was purely accidental, as already noted,
and it was indeed a very
generous and knowledgeable
referee from {\it Phys. Rev. Lett.} who brought Wigner's 1962 work to
my attention (see acknowledgements).
{\it Now} that such a theory (i.e., the theory for a Wigner-type boson)
exists and no  bosons
\footnotemark[4]
\footnotetext[4]{Even though in this presentation we confine our attention
to spin one, the results that we report are valid for bosons of spin one and
greater. See Ref. \cite{BWW} for details.}
 exist (as far as the nature has revealed its
secrets to us so far) that are of Wigner type leads us to the conclusion that
{\it either, in some future experiment (or as a theoretical inevitability
in a much broader theoretical framework),
we shall discover the spin-one Wigner boson that the new theory describes;
or there is some deep underlying reason, yet to be discovered, that
prohibits the existence of the spin-one Wigner boson as a physical reality.}

Assuming that no subtle error of any significance exists in our or
Weinberg's work presented in Refs. \cite{BWW,SWqf,SW}, the above conclusion
is inescapable. Given this situation, we take the liberty of reviewing at
this conference on  ``The Present Status of Quantum Theory of Light: A
Symposium to Honour  Jean-Pierre Vigier'' the theory of the new type of
boson and show that in the massless limit the theory suggests some very
definite and fundamental modifications to Maxwell equations.

\section{A New Type of Spin-One Boson}

Consideration of the space-time symmetries $\Rightarrow$ $(1,0)$
and $(0,1)$
matter fields Lorentz transform (``boost'') in the following fashion:
\footnotemark[5]
\footnotetext[5]
{Our notations and conventions are closest to those found in  Ryder's book
\cite{QFT} on quantum field theory. The
reader not familiar with representations
of the Lorentz group will find a very readable discussion in Chapter 2 of
Ryder's book.}
\begin{eqnarray}
&&(1,0):\quad \phi_R(\vec p\,)\
\,=\,\exp\left(+\,\vec
J\cdot\vec\varphi\,\right)\,\phi_R(\vec 0\,)\quad,\label{r}\\
&&(0,1):\quad\phi_L(\vec p\,)\,
 \,=\,\exp\left(-\,\vec J\cdot\vec\varphi\,
\right)\,\phi_L(\vec 0\,)\quad. \label{l}
\end{eqnarray}
 $\vec J = 3\times3$
angular momentum matrices with $J_z$ diagonal.\\
$\vec \varphi =$  {\it the boost parameter}
defined as
\begin{equation}
\cosh(\varphi\,)
\,=\,{E\over m},\quad\quad \sinh(
\varphi\,)\,=\,{| \vec p\, |\over m},\quad\quad\hat\varphi={\vec  p
\over| \vec p\,|}. \label{bp}
\end{equation}
$\vec p =$ the three-momentum of the particle (of mass m). \\
Note: No ``$i$'' in the argument of exponentials that appear in the above
equations.\\
Reason: $\vec K$, the generator of the boost, $=\pm i\vec J$. The plus sign
for the $(0,1)$-, and minus sign for $(1,0)$-, matter fields. \\
Under parity:  $(1,0) \rightleftharpoons (0,1)$. Parity covariance
$\Rightarrow$
we introduce the $(1,0)\oplus(0,1)$ representation space
spinor
\begin{equation}
\psi(\vec p\,)\,=\,
\left(
\begin{array}{cc}
\phi_R(\vec p\,)\\
\phi_L(\vec p\,)\\
\end{array}\right)\quad.\label{crs}
\end{equation}

As the reader will soon discover, and as I learned from the generalization
\cite{DVAth} of the work of Ryder \cite{QFT},  the  wave
equation satisfied by this $(1,0)\oplus(0,1)$ spinor is determined by the
boost properties of the $\phi_R(\vec p)$ and $\phi_L(\vec p)$, given in
Eqs. (\ref{r}) and (\ref{l}), {\it and} the algebraic relation between
these fields at zero momentum. In the past it had been argued  (see p.
44 of Ryder's book on quantum field theory \cite{QFT}), and it does not
matter  which specific spin in the $(j,0)\oplus(0,j)$ representation space
is under consideration, that ``when a particle is at rest, one cannot
define its spin as either left- or right-handed, so'' for zero momentum
``$\phi_R(\vec 0)=\phi_L(\vec 0)$.'' That this is not true for the
$(1/2,0)\oplus(0,1/2)$ Dirac field is manifest when we look at the explicit
form of zero-momentum spinors.
\footnotemark[6]\footnotetext[6] {This
observation was first made in the Summer of 1991 by C. Burgard \cite{CB}
while trying to understand Ryder's {\it ab initio}
derivation \cite{QFT} of the Dirac equation.}
In canonical representation the argument is made in Ref. \cite{BWW} for the
$(j,0)\oplus(0,j)$ representation space;  and in Weyl (also called chiral)
representation [the representation in which $\psi(\vec p)$ is written in
Eq. (\ref{crs})] the reader should carefully follow Weinberg's arguments on
pp. 220-224 of his text \cite{SWqf} to obtain spin-$1/2$ zero-momentum
spinors (given in Eqs. 5.5.35 and 5.5.36 of \cite{SWqf}) to arrive at the
same conclusion as us for $j=1/2$.

Given Eqs. (\ref{r}) and (\ref{l}), the spinors at momentum $\vec p$,
$\psi(\vec p)$,  are
known if we specify the zero-momentum spinors
$\psi(\vec 0)$. From the work of Ref. \cite{BWW}
and the detailed study on the
closely related subject contained in \cite{SWqf}
it is clear that it is the choice of the zero-momentum spinors,
i.e., specification of $\psi(\vec 0)$, that must
determine the parity, and other related structure, of the theory.
\footnotemark[7]
\footnotetext[7]{This we assert, not as a theorem, but as
an essentially unavoidable conclusion made in an attempt to reconcile the
apparently contradicting conclusions of Refs. \cite{BWW} and \cite{SWqf,SW}.
In principle, of course, there remains the possibility that something
subtle is being missed  by the authors of Ref. \cite{BWW} or/and the author
of \cite{SWqf}.}
In Ref. \cite{BWW} we postulated that in the {\it canonical representation}
\footnotemark[8]
\footnotetext[8]{
Canonical representation is defined as:
\begin{equation}
\psi_{\mbox{canonical}}(\vec p)
={1\over\sqrt 2}
\left(
\begin{array}{cc}
I &I\\
I &-I
\end{array}\right)
\psi(\vec p)
= {1\over\sqrt 2}
\left(\begin{array}{c}
\phi_R(\vec 0) + \phi_L(\vec 0)\\
\phi_R(\vec 0) - \phi_L(\vec 0)
\end{array}\right) .\label{car}
\end{equation}
The $\psi(\vec p)$ in the middle of the above equation refers
to the Weyl representation spinor of Eq. (\ref{crs}) and
$I=3\times 3$ identity matrix.}
the  six-dimensional
$(1,0)\oplus(0,1)$ representation space for a particle at rest can be
spanned by the six zero-momentum spinors:\\
\begin{eqnarray}
\leftline{\mbox{Canonical Representation:}}\nonumber\\
u_{+1}(\vec 0) =\left(\begin{array}{c}
m\\0\\0\\0\\0\\0
\end{array}\right),\quad
u_{0}(\vec 0) =\left(\begin{array}{c}
0\\m\\0\\0\\0\\0
\end{array}\right),\quad
u_{-1}(\vec 0) =\left(\begin{array}{c}
0\\0\\m\\0\\0\\0
\end{array}\right), \label{ur}\\
v_{+1}(\vec 0) =\left(\begin{array}{c}
0\\0\\0\\m\\0\\0
\end{array}\right),\quad
v_{0}(\vec 0) =\left(\begin{array}{c}
0\\0\\0\\0\\m\\0
\end{array}\right),\quad
v_{-1}(\vec 0) =\left(\begin{array}{c}
0\\0\\0\\0\\0\\m
\end{array}\right).\label{vr}
\end{eqnarray}
The $\sigma =0,\pm1$ on $u_\sigma(\vec 0)$ and $v_\sigma(\vec 0)$ carry the
meaning of the projection of spin on the z-axis,
and the possibility of relative phases between various spinors
(not important for the present considerations) are left implicit.
On studying the
C, P, and T properties of the associated wave equation, it turns out that
the $u$- and $v$-spinors are related by the operation of Charge conjugation
and carry opposite relative intrinsic parities \cite{BWW}.

Setting $\vec p=\vec 0$ in Eq. (\ref{car}), of footnote [8],
 and comparing the resulting
equation
with Eqs. (\ref{ur}) and (\ref{vr}) we find that for the $u$-spinors
$\phi_R(\vec 0)= +\phi_L(\vec 0)$
and for the $v$-spinors
$\phi_R(\vec 0)= -\phi_L(\vec 0)$.

To sum up, therefore, we
have the needed algebraic relation between
$\phi_R(\vec p)$ and $\phi_L(\vec
p)$ at zero momentum:
\begin{equation}
\phi_R(\vec 0\,)\, \,=\,\wp_{u,v}\,\phi_L(\vec 0\,)\,,
\label{fr}
\end{equation}
with $ \wp_{u,v}= +1$ for the $u$-spinors and $ \wp_{u,v}= -1$ for the
$v$-spinors.

The three $u_\sigma(\vec p)$ and the three $v_\sigma(\vec p)$
spinors, with $\sigma=0, \pm 1$ representing the three spinorial degrees of
freedom, are obtained by applying the $(1,0)\oplus(0,1)$ boost implicit in
definition (\ref{crs}) and the $(1,0)$ and $(0,1)$ boosts given in Eqs.
(\ref{r}) and (\ref{l}). That is,
\footnotemark[9]
\footnotetext[9]{Note: we stay in the Weyl representation
unless specifically indicated otherwise. For example,
  Eqs. (\ref{ur}) and
(\ref{vr}) are written in canonical representation and this is explicitly
noted right above these expressions.}
\begin{eqnarray}
u_\sigma(\vec p)= \left(
\begin{array}{cc}
\exp(+\vec J\cdot\vec \varphi) & 0\\
0 &\exp(-\vec J\cdot\vec \varphi)
\end{array}\right) u_\sigma(\vec 0)
 ,\\
v_\sigma(\vec p)= \left(
\begin{array}{cc}
\exp(+\vec J\cdot\vec \varphi) & 0\\
0 &\exp(-\vec J\cdot\vec \varphi)
\end{array}\right) v_\sigma(\vec 0); \label{uv}
\end{eqnarray}
with $\sigma=0,\pm 1$.

The reader will note that the six $(1,0)\oplus(0,1)$ spinors thus obtained
follow purely from the projective representations of the Lorentz group and
the associated boosts, and the  $(1,0)\oplus(0,1)$ fields operator for the
quantum description of these particles follows from the canonical arguments
of translational invariance etc. [for details refer to any recent book on
quantum theory of fields, such as \cite{SWqf}]
\begin{equation}
\Psi(x) \,= \,
\sum_{\sigma=0,\pm 1}
\int {d^3p\over (2\pi)^{3} } {1\over 2\,\omega_{\vec p}}
\Big[ u_\sigma(\vec p\,)\, a_\sigma(\vec p\, )\, e^{-i p\cdot x}
+  v_\sigma(\vec p\,) \,b^\dagger_\sigma(\vec p\,) \,e^{i p\cdot x} \Bigr ]
\quad,\label{fo}
\end{equation}
where $\omega_{\vec p\,} = \sqrt{m^2 + {\vec p\,}^2}$; and
$[a_\sigma(\vec p\,),\,a^\dagger_{\sigma'}(\vec p^{\,\prime}]_\pm=
\,\delta_{\sigma\sigma'}
\delta(\vec p-\vec p^{\,\prime})$; etc. and we leave implicit the
possibility that the $b^\dagger$ may carry a hidden numerical factor.

It is now a straightforward mathematical exercise to couple
the right- and left-handed fields to obtain the {\it free-field wave
equation}
for the $(1,0)\oplus(0,1)$ spinors. Using Eq. (\ref{fr}) on the
right-hand side of Eq. (\ref{r}) to re-express $\phi_R(\vec 0)$ in terms of
$\phi_L(\vec 0)$ and then using Eq. (\ref{l}) to replace
$\phi_L(\vec 0)$ by $\exp(\vec J\cdot\vec\varphi)\phi_L(\vec p)$, and
executing a similar exercise beginning with the right-hand side of
Eq. (\ref{l}), we obtain two coupled equations for
$\phi_R(\vec p)$ and $\phi_L(\vec p)$. These two equations are then
transformed into a single wave equation  for the $(1,0)\oplus(0,1)$
spinor (\ref{crs}). This wave equation for the $(1,0)\oplus(0,1)$ spinor
reads:
\footnotemark[10]\footnotetext[10]
{The procedure is in fact valid for {\it any}
spin in the $(j,0)\oplus(0,j)$ representation space.
In particular, when applied for $j=1/2$ one obtains the well-known Dirac
equation.}
\begin{equation}
\left(
\gamma_{\mu\nu}\,p^\mu p^\nu
\,-\,{\wp_{u,v}}\,m^{2} I\right)\,\psi(\vec p\,)\,=\,0\,,\label{eqn}
\end{equation}
with
\begin{eqnarray}
&&\gamma_{\mu\nu}p^\mu p^\nu\,=\,
\left(
\begin{array}{cc}
0&B\,+\,2\,(\vec J\cdot\vec p\,)\,p^0\\
B\,-\,2\,(\vec J\cdot\vec p\,)\,p^0 & 0
\end{array}\right), \label{ga} \\
&&\mbox{where}
\,\,B\,=\,\eta_{\mu\nu}\,p^\mu p^\nu \,+\, 2\,(\vec J\cdot\vec p\,)
\,(\vec J\cdot\vec p\,)
\quad.
\end{eqnarray}
Here, $\eta_{\mu\nu}$ is the
flat space-time metric with the diagonal
$(1,-1,-1,-1)$. The ``$0$'' on the diagonal represents a $3\times
3$ block of zeros. The off-diagonal terms are the $3\times 3$
block matrices.

{}From Eq. (\ref{ga}) we read off the following ``gamma matrices''
(not to be confused with the Dirac ``gamma matrices''):
\begin{eqnarray}
\gamma_{00}=\left(\begin{array}{cc}
0&I\\
I&0
\end{array}\right),
\gamma_{i0}=\gamma_{0i}=\left(\begin{array}{cc}
0&J_i\cr
-J_i&0\end{array}\right),\\
\gamma_{ji}=\gamma_{ij}=
\left(\begin{array}{cc}
0&I\\
I&0
\end{array}\right)\eta_{ij} +
\left(\begin{array}{cc}
0&\{J_i,J_j\}\cr
          \{J_i,J_j\}&0\end{array}\right),
\end{eqnarray}
$i$ and $j$ run over a spacial index $1,2,3$.

For the $(1,0)\oplus(0,1)$ representation space
$\psi(x)\equiv \psi(\vec p\,) \exp(-i\wp_{u,v}\,p\cdot x)$,
and Eq. (\ref{eqn}) requires:
\begin{equation}
\left(
\gamma_{\mu\nu}\partial^\mu\partial^\nu\,+\,
\wp_{u,v}\,m^2\,I
\right)
\,\psi(x) \,=\,0.\label{ea}
\end{equation}

This wave equation is identical to Steven Weinberg's equation for the
$(1,0)\oplus(0,1)$ representation space in all aspects except an important
factor of $\wp_{u,v}$ attached to the mass term. It is this factor that leads
to a fundamentally different, i.e., Wigner type, CPT structure in our theory.
The CPT analysis of the theory is presented in Ref. \cite{BWW} and is found to
be very intricately related to the $\wp_{u,v}$ factor of our theory. The most
important result that emerges from this analysis, and we simply quote it here,
is that {\it the operations of Charge conjugation and Parity anticommute in a
quantum field theory built upon wave equation (\ref{ea}) and field operator
(\ref{fo}) and
thus results in the  underlying spin-one boson and
antiboson carrying  opposite relative intrinsic parities. } To the best of
our knowledge the mathematical construction of Ref. \cite{BWW}, and further
discussed here in its physical content, is the first explicit and
non-trivial example of a quantum theory of the Wigner type. We look forward
to further work on physical and mathematical implications of such a
construct for future fundamental works in quantum field theory and possible
experimental discovery of the Wigner-type bosons.

So far we have essentially emphasized the differences of our construct and
that of Weinberg. We now discuss the aspects that are same (or similar) in
both theories. The dispersion relations associated with both theories are
determined by setting the determinant of $\left( \gamma_{\mu\nu}\,p^\mu p^\nu
\,-\,{\wp_{u,v}}\,m^{2} I\right)$ (for our theory), and the determinant of
$\left( \gamma_{\mu\nu}\,p^\mu p^\nu \,-\,\,m^{2} I\right)$ (for
Weinberg's theory),  equal to zero.
The resulting equation is a 12th-order polynomial in $(E,\vert\vec
p\vert, m)$
and results in the  dispersion relations summarized in Table I.

\begin{table}[htb]
\begin{center}
\caption{Dispersion relations $E=E(p,m)$ associated with Eq.
(17)
and similar equation of Weinberg. $p\equiv\vert\vec p\vert$}
\begin{tabular}{lll}
\hline
  Dispersion Relation &  Multiplicity &  Interpretation \\
\hline
 $E=+\sqrt{p^2 + m^2}$ & $3$ & Causal, ``particle''\cr
                       &   &  $u_{{\pm 1}}(\vec p)$,~
                                       $u_{ {0}}(\vec p)$ \\
 $E=-\sqrt{p^2 + m^2}$ & $3$ & Causal, ``antiparticle''\\
                       &   &  $v_{\pm {1}}(\vec p)$,~
                                       $v_{ {0}}(\vec p)$ \\
 $E=+\sqrt{p^2-m^2}$ & $3$ & Acausal, Tachyonic\\
 $E=-\sqrt{p^2-m^2}$ & $3$ & Acausal, Tachyonic\\
\hline
\end{tabular}
\end{center}
\end{table}
There are two observations that we wish to make in regard to the
tachyonic solutions:
\begin{enumerate}
\item
In the $m\rightarrow0$ limit all dispersion relations are
non-tachyonic.
\item
 For $m\neq 0$, the the tachyonic solutions may be reinterpreted,
on introducing a quartic self-coupling. The ``{\it negative} mass squared''
term can then be interpreted as in the simplest versions of field theories
with broken symmetry. Such an analysis \cite{DVAtg}   shows that the
resulting theory  describes four particles: two charged particles of mass
$m$ (of Wigner type in our theory and usual type in Weinberg's theory), a
(Goldstone-like) spin-one massless (Majorana-like) particle, and a massive
(Majorana-like) spin-one particle of mass $\sqrt 2\vert m\vert$.
\end{enumerate}
In any case, at this stage negative mass squared may be considered
to provide the sought-after physical origin of the  ``$- m^2$'' that appears
in the simplest versions of field theories with broken symmetry.
Or, in absence of this re-interpretation one may
argue that these solutions violate
the original input on the mass parameter via equation (\ref{bp}), and hence
may be considered physically inadmissible (without perhaps interactions
that make it possible to re-interpret the   ``$- m^2$ as above!).
While we do not suspect that usual interactions can induce
transitions between the $E=\pm\sqrt{\vec p^{\,2} + m^2}$ and
$E=\pm\sqrt{\vec p^{\,2} - m^2}$ sectors, we do not know of any proof on the
subject.

\section{The Massless Limit and Maxwell Equations}

That the massless limit is well behaved for the representation space
under consideration was shown in the sixties by Weinberg \cite{SW}.
So to obtain the massless limit of the the theory under consideration,
one may wish to set $m=0$ in
(\ref{eqn}), or equivalently in (\ref{ea}), and argue that
$\eta_{\mu\nu} p^\mu p^\nu \phi_R(\vec p)=0$ and
$\eta_{\mu\nu} p^\mu p^\nu\phi_L(\vec p)=0$ for a massless particle, to
obtain:
\begin{eqnarray}
2\vec J\cdot\vec p\left(\vec J\cdot\vec p
+p^0 I\right) \phi_R(\vec p)=0, \label{mxa}\\
2\vec J\cdot\vec p\left(\vec J\cdot\vec p
-p^0 I\right) \phi_L(\vec p)=0; \label{mxb}
\end{eqnarray}
and since it is  known (see, for example, \cite{DVAth}) that
\begin{eqnarray}
\left(\vec J\cdot\vec p+p^0 I\right) \phi_R(\vec p)=0 \label{ma}\\
\left(\vec J\cdot\vec p-p^0 I\right) \phi_L(\vec p)=0, \label{mb}
\end{eqnarray}
are indeed free Maxwell equations one  may claim that we have obtained
Maxwell equations in the limit $m=0$ of our, or that of Weinberg's, theory.
{\bf No}, this is not so. The reason, as we already noted a few years ago
\cite{DVAm}, is that Maxwell  equations (\ref{ma}) and (\ref{mb}) do {\it
not} follow from Equations (\ref{mxa}) and (\ref{mxb}) {\it because} the
matrix $2\vec J\cdot\vec p$ is {\it non-invertible}. The Determinant $(2\vec
J\cdot\vec p)$ identically vanishes.

The analysis of the massless limit is not yet complete, but at this stage
one can already note that all solutions of (Maxwell)  Eqs. (\ref{ma}) and
(\ref{mb}) are solutions of Eqs. (\ref{mxa}) and (\ref{mxb}), but there may
exist solutions that do not satisfy Maxwell equations but still are
solutions of Eqs. (\ref{mxa}) and (\ref{mxb}). Another point to note is
that   (Maxwell) Eqs. (\ref{ma}) and (\ref{mb}) are first order in
space-time derivatives.  Eqs. (\ref{mxa}) and (\ref{mxb}) are of second order
in space-time derivatives. Hence at least {\it additional  boundary conditions
are to be satisfied. Any departures, therefore, from Maxwell equations will
only be expected, or are most likely,  for phenomenon that involve strong
fields and/or strongly varying fields.}
\footnotemark[11]\footnotetext[11]{What sets the scale that determines
``strong'' in the above statement? This question requires a precise answer,
and we wish to take up this subject in the future. But for the moment we shall
assume that a definite scale can be defined, or at least that the question
can be answered in a specific experimental set up. }

To complete the story we note that in the beginning of this section we assumed
``$\eta_{\mu\nu} p^\mu p^\nu \phi_R(\vec p)=0$ and $\eta_{\mu\nu} p^\mu
p^\nu\phi_L(\vec p)=0$ for a massless particle.'' There is no justification
to invoke this assumption {\it a priori}. Therefore, the  Maxwell equations
(\ref{ma}) and  (\ref{mb}) should, rigorously speaking,
be replaced, instead of (\ref{mxa}) and
(\ref{mxb}), by:
\begin{eqnarray}
\left(\eta_{\mu_\nu} p^\mu p^\nu I + 2\vec J\cdot\vec p\left(
\vec J\cdot\vec p + p^0 I\right)\right)\phi_R(\vec p)=0,\label{ema}\\
\left(\eta_{\mu_\nu} p^\mu p^\nu I + 2\vec J\cdot\vec p\left(
\vec J\cdot\vec p - p^0 I\right)\right)\phi_L(\vec p)=0.\label{emb}
\end{eqnarray}
In reference to the above equations and Eqs. (\ref{mxa}),
(\ref{mxb}), (\ref{ma}), and (\ref{mb}),
it seems important to observe that these equations have solutions
only if the appropriate dispersion-relation  determining determinant vanishes.
These determinants are:
\begin{eqnarray}
&&\mbox{Determinant}
\left(
\vec J\cdot\vec p \pm p^0 I\right) = \mp E\left(\vec p^{\,2}-E^2\right) \\
&&\mbox{Determinant} \left(2\vec J\cdot\vec p\left(
\vec J\cdot\vec p \pm p^0 I\right) \right)=(\mbox{identically})\,\, 0,\\
&&\mbox{Determinant}
\left(\eta_{\mu_\nu} p^\mu p^\nu I + 2\vec J\cdot\vec p\left(
\vec J\cdot\vec p \pm p^0 I\right)\right) =
-(\vec p^{\,2} -E^2)^3 .
\end{eqnarray}
A comparison of the dispersion relations implied by setting the above
determinants to zero with the dispersion relations for $m\neq 0$ case,
tabulated in Table I, again indicates Eqs. (\ref{ema}) and (\ref{emb}) as
the sole candidate for the massless limit of our (or, Weinberg's) theory.

So, to sum up our analysis of the massless limit of our (or that of
Weinberg's theory), we conclude that: {\it Present theoretical arguments
suggest that in strong fields, or high-frequency phenomenon, Maxwell
equations may not be an adequate description of nature. Whether this is so
can only be decided by experiment(s). Similar conclusions, in an apparently
very different framework, have been independently arrived at by M. Evans
\cite{MyronE} and communicated to the author.  }

\section{Concluding Remarks}

First, we showed that in the $(1,0)\oplus(0,1)$ representation space there
exist not one but  two theories for charged particles.
\footnotemark[12]\footnotetext[12]{Considerations of particles that are
self-charge conjugate leads to a yet another theory
in the $(1,0)\oplus(0,1)$ representation space. This construction appears
 in Ref. \cite{DVAn}.} In the Weinberg construct, the boson and its
antiboson carry same relative intrinsic parity, whereas in our construct
the relative intrinsic parities of the boson and its antiboson are
opposite. These results originate from the commutativity of the operations
of Charge conjugation and Parity in Weinberg's theory, and from the
anti-commutativity of the operations of Charge conjugation and Parity in
our theory. We thus claim that we have constructed a first non-trivial
quantum theory of fields for the Wigner-type particles.
\footnotemark[13]
\footnotetext[13]
{Despite the fact that the $\wp_{u,v}$ factor appears only in the mass term
one cannot claim that Weinberg's theory and our theory have the same
physical content in the massless limit. The differences may arise from how
the zero momentum $(1,0)\oplus(0,1)$ spinors are chosen in the two
theories. This certainly is true for the differences in the two theories
for massive particles. This aspect requires further study. Rigorously
speaking, the ``rest spinors'' and ``zero-momentum spinors'' must be
distinguished while speaking of the massless limit. For  massless
particles there are no rest spinors.}
Second, the massless limit of both theories seems formally identical  and
suggests a fundamental modification of Maxwell equations.
At its simplest level, the modification to Maxwell equations enters via
additional boundary condition(s).

\acknowledgements

This an unusually long acknowledgement for an anonymous referee who has
inspired and taught me in many ways. So I begin with a few personal
comments in the nature of  an introduction. I am deeply aware of the
unusual nature of this acknowledgement and the criticism that it may draw,
but to keep its contents in my files will be unfair to my fellow students
of physics for many reasons.

In part, the purpose of this long addenda in the form of an acknowledgement
to the manuscript  is to  document and acknowledge  the significant
contributions of this anonymous referee from {\it Phys. Rev. Lett.} in the
construction of the Wigner-type quantum theory of fields. That the
construction of a Wigner-type quantum theory of fields was purely
accidental will also become apparent in the process of reading the two
reports (from the same referee) that follow. To fully appreciate the impact
of the referee on our work the reader should read Ref. \cite{BWW}
concurrently. While the physics these reports contain is important, it is
equally important how that physics weaves with history and personal
affection. Having talked about phases, projective representations, and
given proofs of some important theorems, and having asked many questions,
the referee suddenly seems overtaken by his affection for the man from whom
he must have learned much of all this and writes ``I am not professor
Wigner (he has been 90 this month; let him in peace).''

I treat these reports as little monographs and these little monographs have
provided me much guidance in the construction of the theory that I reviewed
here. These reports are exceptional in their clarity, unsurpassed in their
generosity, and exhibit a deep affection that their author holds for
physics and the giants in his field. I reproduce these reports here not
only to document the contributions that this referee made to my work, but
also in the hope that future generations will be inspired and guided by the
content and style of these reports when they write their reviews for the
manuscripts of their colleagues. I remain deeply thankful to so unusual a
referee. The very existence of this referee gives me hope for the future of
physics in these difficult cultural times when so little of support exists
for fundamental science and it is so difficult to find a true mentor in
the classic sense of this word. For the roughly six-month period during
which I worked on the revision of the manuscript the referee became
my mentor, and perhaps my collaborators too gained from the knowledge
and good advice of the referee.

The text of the two reports that appears here is the unedited exact
version, and all spellings are left as in the original to preserve complete
flavour of the original reports. For example at one place we have
``Majorana articles'' (instead of Majorana particles); at other particles
appears as ``particules.'' Similarly all punctuation, spacing after commas,
italicized letters and boldface letters are reproduced as they appear in
the original.
\footnotemark[14]\footnotetext[14]
{
No other referee report was received by the editors of {\it Phys. Rev.
Lett.} and despite a strong recommendation  from the referee to publish the
paper the manuscript was rejected by the editors. The manuscript, with the
sole addition of an acknowledgement to the anonymous referee, was then
submitted to {\it Phys. Lett. B} where it was independently reviewed and
accepted without any revisions.
}

\noindent
{\bf First Report:}

{\it This paper cannot be published in any physical journal.} Indeed it
presents an interesting idea but this idea is not new: for instance
the possibility to have boson  anti-boson
relative parity is presented (with an equivalent, but different
representation of the Lorentz group for particle states) in the
E.P. Wigner (the physicist who has introduced parity in quantum mechanics
in 1928) contribution to the volume:
{\it Group Theoretical concepts and methods in Elementary particle
Physics.}
F. G\"ursey editor, Gordon and Breach, New-York 1964.

In this paper Wigner shows that to have opposite relative parity
for boson anti-boson one has to pay a price: the doubling of the number of
states, i.e. a new type of degeneracy (referred to as ``Wigner type'' by
the knowledgeable physicists). So the author should refer to this Wigner
paper and since he does not claim a doubling of the usual number of states
he has to explain where is the discrepancy with Wigner. If he succeeds that
will be very interesting to publish it. Unhappily there is not enough
information in the manuscript to know if the author may succeed. For
instance the author should write explicitly the Charge Conjugation operator
$U(C)$ in his field theory.

Here are comments and questions for helping the author:

1) In the traditional notation $(j,0)\oplus(0,j)$ for the finite
dimensional reducible representation of the connected Lorentz group, the
parity operation exchanges these inequivalent irreducible components. So to
use equation (2) (taken from your thesis) you have to give more precisions
on what
you have done: the parity operator cannot be block diagonal in the
direct sum;
probably you have already taken a symmetric and antisymmetric combinations
of these two subspaces which carry inequivalent representations of the
connected Lorentz group? The reader needs to know in detail. Redactional
details: your equation
(3) contains no information as long as you do not give $t', x'$ as function
of $t,x$ ! Why do you call spinors the vectors $(1,0$ and $(0,1)$ which
correspond to $\vec E\pm i\vec B$ in electromagnetism (they were
introduced by Helmoltz).

2) You do quote Wigner paper of 1939 on the inhomogenuous Lorentz group,
but it is irrelevant here (e.g. in this beautiful and very important
mathematical paper, Wigner studied the unitary representations of the full
group, including time reversal, although he did  know that in physics
time reversal has to be represented by an antiunitary operator: see his
paper of 1932 in G\"ottingen Nachrichten). The important fact to remember
from
this 1939 paper (or from Wigner's earlier book) is that in quantum
physics, the relativity group is realized by a projective representation,
i.e. a representation up to a factor. The phase of the unitary operator
representing a discrete operation is therefore arbitrary; and you are
completly right to emphasize after your equation (6) that there
is a ``{\it
global} phase factor'': to neglect it, as you suggest, might be throwing
the baby out with the water of the bath tub. The way out is simple:
independently of the arbitrary phase of the parity operator,
{\bf the relative phase $\epsilon$ between particle and anti particle
states is not arbitrary and is naturally defined as:}
\[
U(P) U(C) =\epsilon U(C) U(P) \quad\quad\quad\quad (1)
\]
It is a simple exercise to prove $\epsilon^2 =1$ by using associativity of
the group law. Indeed
multiplying (1) on the right you obtain $U(P) U(C)^2=\epsilon U(C) U(P)
U(C)$
and using (1) again $U(P) U(C)^2 = \epsilon^2 U(C)^2 U(P) =\epsilon^2 U(P)
U(C)^2$.

In your theory, you should define explicilty the charge conjugation
operator $U(C)$.
Does it commute or anticommute with $U(P)$?

3) In the frame of quantum field theory do not forget that operators which
correspond to physical identity, as for instance those representing
$C^2$, $P^2$, $(CP)^2$, $T^2$, $(CPT)^2$ etc... are not a multiple of the
identity operator, but their value depends on the superselection sector on
which they act. The values of the square of the antiunitary operators
are well fixed
on each superselection sector; (this is not the case of the unitary
operators such as $P^2$ for the non vacuum sectors, but the relative
value between different sectors might be
well defined). The proof uses again the associativity of the group law. Let
$V=U K$ be an antiunitary operator: in a basis it is the product of a
unitary operator and of the complex conjugation $K$. We assume that the
restriction to a superselection sector of the square of this antiunitary
operator is a multiple of the identity: $V^2=\omega I$; since $V^2$ is
unitary, $\omega$ is a phase: $\overline\omega\omega=1$. Then $V^3=\omega
V=V\omega=\overline\omega V$; so $\omega=\overline\omega=\pm1$. For
instance, in usual
quantum field theory, without the ``Wigner type'' degeneracy, $V(CPT)^2 =1$
on the bosonic sector and $-1$ on the fermionic sector. This implies that
$V(CPT)^2\psi=-\psi V (CPT)^2$ for any fermion field $\psi$.

4) There is  arbitrariness in defining parity of a single field without
interaction, but this arbitrariness can be completely reduced with the
interactions. Indeed to measure the parity of a particle you must interact
(by a parity conserving interaction!) with the particle,
either to produce it or by the study of its spontaneous decay (the case
interesting for you?).
Do you wish that these particules have electromagnetic interactions: write
their electric current if they have an electric charge. If not they can
still interact by an induced magnetic moment and, for spin 1 particle, a
quadrupole moment. Or do you expect these mesons with absolutly no
electromagnetic interaction?

5) Finally a serious study will lead you to rise the fundamental question
of $CTP$. What do you expect about the $CTP$ behaviour of your particles?
That is the type of question you must be able to answer if you want to make
an interesting contribution to present particle physics. Indeed if you want
their interaction to violate $CTP$ invariance, you have to reconstruct all
present day physics!

I am not professor Wigner (he has been 90 this month; let him in peace).
But I wished to help you; that is why I wrote this long report asking you
many
questions.

\noindent
P.S. While writing this report I have no access to a collection of Physical
Reviews of the sixties. In one of them, if my memory is faithful, you will
find a paper by T.D. Lee and G.C. Wick which deals with the discrete
symmetries $C$, $P$, $T$ in quantum field theory and give the values of the
square of their representing operators on different superselection rules
sectors.

\noindent
{\bf Second Report}

The first named author has appreciated my exceptionally long report.
He has read and well assimilated the literature I suggested.
Congratulations!

This very new version of the manuscript has now three authors and carries a
very well chosen title. Indeed Bargmann, Wightman and Wigner had studied,
this subject forty years ago, in an unpublished book (several chapters were
distributed as preprints). The authors explain well the scope of their
paper. They have made a thorough construction of the field theory of a non
usual Wigner type; that is completely new and all given references are
relevant.
{\it This paper should be published.}

However the authors have missed an important point: in quantum theories,
symmetry groups are implemented through {\bf projective} representations.
As the authors rightly write
(page 7 and footnote 2) one can ignore an overall phase factor
(except in Majorana theory), but a crucial argument of parity should not be
obscured by conventions. In my report, I recalled the {\bf proof}
that for the operators $C$, $P$, $T$ and their products, there are signs of
$\epsilon$ independent from conventions: on each superselection rule
sector,{\it
the value $\epsilon I$ of the square of the antiunitary operators:
$T^2$, $(PT)^2$, $(CT)^2$, $(CPT)^2$ and the sign in $PC=\epsilon
CP$} which indicates the commutation or the anticommutation of $P$ and $C$;
of course,
these last two possibilities correspond respectively to same or opposite
parity for particles and
antiparticles. To stress this important point, the authors should show the
anticommutation of $P$ and $C$ in the theory they develop and compute the
value of the squares of the four antiunitary operators on the bosonic and
fermionic states. Those characterize the Wigner's types.

In his letter presenting this new version, the first named author
points out that for zero mass particles the situation is different
(specialy for Majorana articles). He is completly right and I am looking
forward for the announced paper. I want to bring to his attention that this
was understood in Tiomno thesis (princeton, around 1947-48), unpublished I
believe. Either Tiomno (in Brazil) or Wightman, in Princeton, could give him
more details.

\end{document}